\documentclass[aps,pre,epsf,twocolumn,floatfix]{revtex4-2}

\usepackage{amsmath}
\usepackage{amssymb}
\usepackage{epsfig}
\usepackage{graphicx}
\usepackage[T1]{fontenc}
\usepackage[utf8]{inputenc}
\usepackage{color}
\usepackage[normalem]{ulem}
\usepackage{tikz}
\usetikzlibrary{positioning}

\begin{document} 

\title{Tricriticality and finite-size scaling in the triangular Blume-Capel ferromagnet}

\author{Dimitrios Mataragkas}

\author{Alexandros Vasilopoulos}

\author{Nikolaos G. Fytas}
\email{nikolaos.fytas@essex.ac.uk}

\affiliation{School of Mathematics, Statistics and Actuarial Science, University of Essex, Colchester CO4 3SQ, United Kingdom}

\author{Dong-Hee Kim}
\email{dongheekim@gist.ac.kr}

\affiliation{Department of Physics and Photon Science, Gwangju Institute of Science and Technology, Gwangju 61005, Korea}

\date{\today}

\begin{abstract}
We report on numerical simulations of the two-dimensional spin-$1$ Blume-Capel ferromagnet embedded in a triangular lattice. Utilizing a range of Monte Carlo and finite-size scaling techniques, we explore several critical aspects along the crystal field--temperature ($\Delta, T$) transition line. Wang-Landau simulations measuring the joint density of states in combination with the method of field mixing allow us to probe the phase coexistence curve in high resolution, determining the tricritical point $(\Delta_{\rm t}, T_{\rm t})$ with improved accuracy and verifying the tricritical exponents. Extensive multicanonical simulations identifying transition points across the phase diagram characterize the Ising universality class for $\Delta < \Delta_{\rm t}$ with precise determination of thermal and magnetic critical exponents expected in the second-order regime. On the other hand, for $\Delta > \Delta_{\rm t}$, a finite-size scaling analysis is dedicated to revealing the first-order signature in the surface tension that linearly increases upon lowering the temperature deeper into the first-order transition regime. Finally, a comprehensive picture of the phase diagram for the model is presented, collecting transition points obtained from the combined numerical approach in this study and previous estimates in the literature.
\end{abstract}

\maketitle

\section{Introduction}
\label{sec:intro}

The Blume-Capel model is defined by a spin-$1$ Ising Hamiltonian with a single-ion uniaxial crystal-field anisotropy~\cite{blume,capel}. The fact that it has been very widely studied in statistical and condensed-matter physics is explained not only by its relative simplicity and the fundamental theoretical interest arising from the richness of its phase diagram, but also by a number of different physical realizations of variants of the model, ranging from multi-component fluids to ternary
alloys and $^{3}$He--$^{4}$He mixtures~\cite{lawrie}. In recent years, the Blume-Capel model was invoked in order to understand properties of ferrimagnets~\cite{selke-10}, wetting and interfacial adsorption~\cite{fytas13}, but also the scaling of the zeros of both the partition function~\cite{leila} and the energy probability distribution~\cite{macedo24}.

The zero-field model is described by the Hamiltonian
\begin{equation}\label{eqHamiltonian}
  \mathcal{H}
  =-J\sum_{\langle xy\rangle}\sigma_{x}\sigma_{y}+\Delta\sum_{x}\sigma_{x}^{2}
  = J E_{J}+\Delta E_{\Delta},
\end{equation}
where the spin variables $\sigma_{x}$ take on the values $-1, 0$, or $+1$, $\langle xy\rangle$ indicates summation over nearest neighbors, and $J>0$ is the ferromagnetic exchange interaction. The parameter $\Delta$ is known as the crystal-field coupling and it controls the density of vacancies ($\sigma_{x}=0$). For $\Delta\rightarrow -\infty$, vacancies are suppressed and the model becomes equivalent to the simple Ising ferromagnet. We point out that the decomposition on the right-hand side of Eq.~\eqref{eqHamiltonian} into the bond-related and crystal-field-related energy contributions $E_J$ and $E_\Delta$, respectively, will be useful in the context of the numerical methods discussed below.

\begin{figure}
    \includegraphics[width=1.0\linewidth]{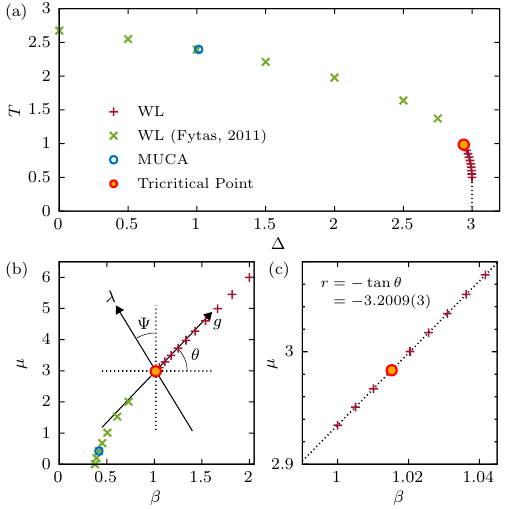}
    \caption{Phase diagram of the two-dimensional spin-$1$ triangular Blume-Capel ferromagnet. (a) Transition points in the plane of the crystal-field coupling $\Delta$ and temperature $T$. The first-order transition line and the location of the tricritical point are determined from the field-mixing analysis using the Wang-Landau (WL) joint density of states. The critical points in the second-order transition regime are computed through multicanonical (MUCA) calculations in this work. For completeness, the earlier Wang-Landau simulation results from Ref.~\cite{fytas_BC} are also presented. We note that the first-order transition points from the multicanonical calculations are omitted, as they are virtually identical to the Wang-Landau results. The complete list of the transition points is given in Table~\ref{tab:delta_c}. (b) The phase diagram is replotted in the $\beta = 1/T$ and $\mu = \Delta/T$ plane to illustrate the two scaling fields, $g$ and $\lambda$, from the field-mixing analysis described in Sec.~\ref{sec:field-mixing}. (c) An enlargement of the phase boundary near the tricritical point is used to estimate the field-mixing parameter $r$ from its slope. \label{fig:phase-diagram}}
\end{figure}

The phase boundary of the model in the crystal field--temperature ($\Delta, T$) plane separates the ferromagnetic from the paramagnetic phase. The ferromagnetic phase is characterized by an ordered alignment of $\pm 1$ spins. The paramagnetic phase, on the other hand, can be either a completely disordered arrangement at high temperature or a $\pm1$-spin gas in a $0$-spin dominated environment for low temperatures and high crystal fields. At high temperatures and low crystal fields, the ferromagnetic--paramagnetic transition is a continuous phase transition in the Ising universality class, whereas at low temperatures and high crystal fields, the transition is of first order~\cite{blume,capel}. The model is
thus a paradigmatic example of a system with a tricritical point $(\Delta_{\rm t}, T_{\rm t})$~\cite{lawrie}, where the two segments of the phase boundary meet. At zero temperature, it is clear that ferromagnetic order must prevail if its energy $zJ/2$ per spin (where $z$ is the coordination number) exceeds that of
the penalty $\Delta$ for having all spins in the $\pm 1$ state. Hence the point $(\Delta_0 = zJ/2, T = 0)$ is on the phase boundary~\cite{capel}. On the other hand, for zero crystal fields, the transition temperature is not exactly known, but it has been well-studied for a number of lattice geometries. For a picture of the phase diagram based on the data of the current work but also on some previous computations from Ref.~\cite{fytas_BC}, we refer the reader to Fig.~\ref{fig:phase-diagram}. For brevity, we set $J = 1$ and $J/k_\mathrm{B} = 1$ as the unit of energy and temperature hereafter.

Since its original formulation, the model~\eqref{eqHamiltonian} has been studied in mean-field theory as well as in perturbative expansions and numerical simulations for a range of lattices, mainly in two and three dimensions, see, e.g.,
Refs.~\cite{fytas_BC,fytas2012,fytas2013,zierenberg2015}. Most work has been devoted to the two-dimensional model on the square lattice, employing a wide range of methods, including real-space renormalization~\cite{berker1976rg}, Monte Carlo simulations, Monte Carlo renormalization-group
calculations~\cite{landau1972,kaufman1981,selke1983,selke1984,landau1986,xavier1998,deng2005,silva2006,hurt2007,malakis1,malakis2,kwak2015},
$\epsilon$ expansions~\cite{stephen1973,chang1974,tuthill1975,wegner1975}, high- and low-temperature series expansions~\cite{fox1973,camp1975,burkhardt1976}, and 
transfer matrix calculations~\cite{beale1986,kim17,blote2019}. 

In this paper we are interested in the critical properties of the Blume-Capel model embedded in the triangular lattice, where $z=6$ and $\Delta_{0} = 3$. For this particular case, Mahan and Girvin~\cite{mahan} were the first to apply position-space renormalization-group methods to solve the model. These authors estimated the critical
frontier with the location of the tricritical point
$(\Delta_{\rm t}, T_{\rm t}) = (2.686, 1.493)$. Many years later, Du \textit{et al.},~\cite{du} performed a further sophisticated analytical calculation using an expanded Bethe-Peierls approximation to find
respectively the estimates $(\Delta_{\rm t}, T_{\rm t})=(2.841, 1.403)$~\cite{comment1}. The refinement $(\Delta_{\rm t}, T_{\rm t})=(2.925(8), 1.025(10))$ was proposed in Ref.~\cite{fytas_BC} via Monte Carlo simulations and an approximate scheme that takes advantage of the scaling behavior of the specific heat. As a historical note, we recall that in the 1980s and 1990s, the spin-$1$ Ising model on the triangular lattice was utilized as a $3$-state lattice-gas model, drawing inspiration from experiments on electrolyte adsorption on single-crystal (111) surfaces of FCC metals. While experimental interest primarily focused on the antiferromagnetic case, comprehensive phase diagrams for both ferromagnetic and antiferromagnetic models, highlighting topologically distinct ground states, can be found in Refs~\cite{rikvold88,collins88}. For a  more recent study of the triangular antiferromagnetic Blume-Capel model we refer the reader to Ref.~\cite{zukovic13}.

Here we utilize a combined and robust numerical protocol based on the Wang-Landau and multicanonical algorithms, and corroborated by the Metropolis scheme, that facilitates the elucidation of several critical aspects of the triangular Blume-Capel ferromagnet. We propose a high-accuracy determination/characterization of the tricritical point using the method of field mixing and we provide a set of phase-transition temperatures mainly in the first-order transition regime. We analyze the scaling behavior of several observables in this first-order transition regime, focusing on the energy probability density functions, through which the scaling behavior of the surface tension is scrutinized upon lowering the temperature. Finally, we verify the expected Ising universality along the second-order transition line by computing the thermal and magnetic critical exponents of the ferromagnetic-paramagnetic transition and showcasing the logarithmic scaling behavior of the specific heat.

The remainder of this paper is organized as follows: In Sec.~\ref{sec:tricriticality}, we present our findings on the location of the tricritical point and the proposed tricritical exponents, utilizing a field-mixing analysis based on Wang-Landau calculations of the joint density of states. Section~\ref{sec:transitions} discusses the transition features and provides a detailed finite-size scaling analysis along both segments of the phase boundary, supported by dedicated multicanonical simulations. Finally, Sec.~\ref{sec:discussion} offers a critical evaluation of our main results alongside a comprehensive overview of the model's phase diagram -- with a focus on the first-order transition regime leading up to the tricritical point -- and outlines directions for future research.

\section{Tricriticality}
\label{sec:tricriticality}

In this section, we identify the tricritical point and validate several conjectured tricritical exponents through a field-mixing analysis. This analysis is based on the joint probability distribution of $E_J$ and $E_\Delta$, which is calculated using the Wang-Landau algorithm in conjunction with histogram reweighting and conventional Monte Carlo methods. Below, we outline the numerical details of computing the joint probability distribution and explain how it is utilized in the field-mixing method to delineate the phase boundary curve and characterize the tricritical behavior.

\begin{figure}
    \includegraphics[width=1.0\linewidth]{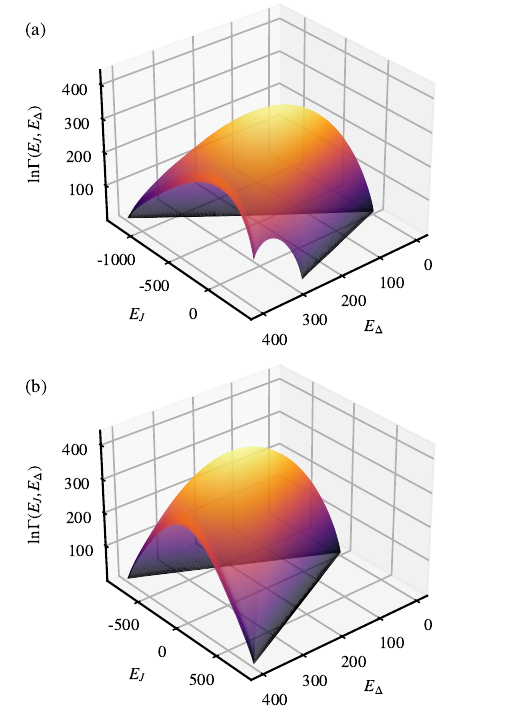}
    \caption{A comparison of the logarithm of the joint density of states, $\ln \Gamma(E_J, E_\Delta)$, for the triangular (a) and square-lattice (b) Blume-Capel models, as obtained through Wang-Landau simulations for a system with linear size $L = 20$.}
    \label{fig:JDOS}
\end{figure}

\subsection{Wang-Landau joint density of states}
\label{sec:WL}

We employed the standard Wang-Landau algorithm~\cite{WL1,WL2} in triangular lattices and periodic boundary conditions with a single walker on the space of $E_J$ and $E_\Delta$ to measure the joint density of states $\Gamma(E_J, E_\Delta)$. Hereafter, $N = L \times L$ denotes the total number of spins and $L$ the linear dimension of the triangular lattice. While the current simulations maintain the same numerical framework as the earlier calculations conducted on square lattices by one of the authors~\cite{kwak2015}, we have adjusted the stopping criterion for the modification factor $f$ and the histogram flatness to accommodate the slower convergence observed in triangular lattices. The flatness criterion for the histogram to update $f$ was set as $95 \%$ for $L \le 14$, $90\%$ for $L = 15$, and $80\%$ for $L > 15$. The stopping criterion of $f$ was set as $10^{-8}$ for $L \le 14$ and $10^{-7}$ for $L > 14$.

Figure~\ref{fig:JDOS} compares the landscape of $\Gamma(E_J, E_\Delta)$ for the triangular and square-lattice systems, revealing a spiky structure at high $E_J$ in the triangular-lattice case that is absent in the square-lattice system~\cite{silva2006,kwak2015}. This spiky structure in the triangular-lattice system is linked to configurations exhibiting frustration. Although the frustration is specific to the antiferromagnetic model, the structure in $\Gamma$ influences the Wang-Landau process regardless of the sign of $J$, as the algorithm must explore the entire energy space, which can slow down the sampling dynamics necessary to satisfy the flatness criterion. Here, Wang-Landau data of the joint density of states were obtained for linear sizes $L = \{8, 9, 10, 12, 14, 15, 16, 18, 20, 21, 22, 24\}$. Five (two) samples of the density of states were prepared from independent simulations for $L < 20$ ($L\ge 20)$. The deviations between different samples turned out invisibly small for the data analyzed in the following sections. In computational time cost, the largest system size $L = 24$ of our current simulations in the triangular model is comparable to the previous calculations conducted on the square-lattice model with $L = 48$~\cite{kwak2015}.

Once the joint density of states is obtained, one can generate the probability distribution of an energy-based quantity $X_{E_J, E_\Delta}$ at a given set of $(\Delta, T)$ without additional measurements simply by enumerating
\begin{equation}
\label{eq:P(X)}
P(X) = \sum_{E_J,  E_\Delta} P(E_J, E_\Delta;\beta, \mu) \delta \left(X - X_{E_J, E_\Delta} \right).
\end{equation}
Note that $P(E_J, E_\Delta;\beta, \mu)$ at any $\beta \equiv 1/T$ and $\mu \equiv \Delta / T$ is directly given by the density of states $\Gamma$ via
\begin{equation}
\label{eq:P_all}
P(E_J, E_\Delta; \beta, \mu) = \frac{1}{\mathcal{Z}} \Gamma(E_J, E_\Delta) e^{-\beta E_J -\mu E_\Delta}.
\end{equation}
In evaluating the probability density $P(X)$, we approximate the discrete
spectrum of the $\delta$-function by Gaussian peaks with small finite broadening.

\subsection{Histogram reweighting}
\label{sec:reweighting}

Despite the advantage provided by the Wang-Landau joint density of states, we should note that it is only available for relatively small system sizes. To solidify our estimate of the tricritical point identified by the Wang-Landau data, we additionally considered the histogram reweighting (HR) technique~\cite{FS1,FS2} based on conventional Monte Carlo calculations of Metropolis type at a specific parameter for larger systems. The histogram $H(E_J,E_\Delta;\beta_0,\mu_0)$ measured at fixed $\beta_0$ and $\mu_0$ can be reweighted to generate the probability distribution of $E_J$ and $E_\Delta$ at nearby $\beta$ and $\mu$ values via
\begin{equation}
\label{eq:reweight}
    P(E_J, E_\Delta; \beta, \mu) \propto H(E_J,E_\Delta;\beta_0,\mu_0)
    e^{-\beta^\prime E_J - \mu^\prime E_\Delta},
\end{equation}
where $\beta^\prime = \beta - \beta_0$ and $\mu^\prime = \mu - \mu_0$. The range of validity in $\beta^\prime$ and $\mu^\prime$ is generally guided by the response function, decreasing as it approaches the transition point and as with increasing system size. To keep the reweighting range as narrow as possible for reliability, we collected the data of histograms at each pseudo-transition point extrapolated from the Wang-Landau estimates for $T \in \{0.97, 975, 0.98, 0.985, 0.99, 0.995\}$. We mostly used the single-histogram reweighting along the $\mu$ axis for a given temperature, while we employed the multiple-histogram variant for the finite-size scaling analysis of the specific heat. Histograms were obtained through $10^6 \times L^2$ Monte Carlo measurements that are split into $100$ bins for the jackknife estimator of statistical errors. The statistical errors turned out to be much smaller than the marker size and the line thickness of the data shown below in Sec.~\ref{sec:TCP}.  

\subsection{Method of field mixing}
\label{sec:field-mixing}

The partition function of the Blume-Capel model at zero magnetic field can be written as
\begin{equation}
\label{eq:Z}
    \mathcal{Z}(\beta, \mu) = \sum_{E_J, E_\Delta} \Gamma(E_J, E_\Delta) e^{(-\beta E_J - \mu E_\Delta)}, 
\end{equation}
where $\beta \equiv 1/T$ and $\mu \equiv \Delta/T$.
As it is well established for Ising critical phenomena in fluid models and the spin-$1$ Blume-Capel model on the square lattice~\cite{wilding92,bruce92,wilding96,plascak13,silva2006,kwak2015}, the phase transition in the $\beta-\mu$ plane is asymmetric in the vicinity of the tricritical point, where $\beta$ and $\mu$ are not the relevant scaling fields. Instead, the relevant scaling fields are written as their linear combinations
\begin{eqnarray}
    \lambda &=& (\mu - \mu_{\rm t}) + r (\beta - \beta_{\rm t}), \\
    g &=& (\beta - \beta_{\rm t}) + s (\mu - \mu_{\rm t}), 
\end{eqnarray}
where $s$ and $r$ are the system size-dependent field-mixing parameters, and $\beta_{\rm t}$ and $\mu_{\rm t}$ indicate the tricritical point. The direction of the scaling field $g$ is tangent to the phase-coexistence curve. The field $\lambda$ is across the phase boundary but is not necessarily orthogonal to $g$. The directions of $\lambda$ and $g$ are graphically described in Fig.~\ref{fig:phase-diagram}(b), where the field-mixing parameters are associated with the angles as $r = -\tan\theta$ and $s = \tan\Psi$. The scaling operators conjugate to $\lambda$ and $g$ are 
\begin{eqnarray}
    \mathcal{Q} &=& \frac{1}{1-rs}(n - s \epsilon), \label{eq:Q}\\ 
    \mathcal{E} &=& \frac{1}{1-rs}(\epsilon - rn), \label{eq:E}
\end{eqnarray}
respectively, where $\epsilon \equiv L^{-d} E_J$ and $n \equiv L^{-d} E_\Delta$ with dimension $d = 2$ in our model, satisfying the relation $\langle\mathcal{O}_x\rangle = -L^{-d} \partial_x \ln \mathcal{Z}$. 
Note that $n$ represents the density of non-zero spin states, and the operator $\mathcal{Q}$ deviates from $n$ when $s\neq 0$. For example, at $T = 0.985$, our estimates are $s \approx 0.028$ for $L = 24$ and $s \approx 0.085$ for $L = 48$, as we discuss in the analysis below. The probability distribution of the operator $\mathcal{Q}$ is a crucial tool in the field-mixing analysis, enabling the determination of the phase-coexistence curve and the identification of the tricritical point.

\begin{figure}
    \includegraphics[width=1.0\linewidth]{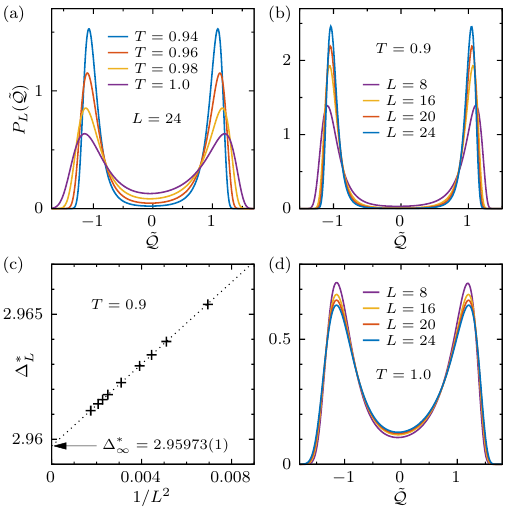}
    \caption{Locating a transition using the field-mixing method and Wang-Landau simulations: (a) Probability distribution $P_L(\tilde{\mathcal{Q}})$ of the scaling variable $\mathcal{Q}$ determined for $L = 24$ along the phase transition curve. The system-size dependence of $P_L(\tilde{\mathcal{Q}})$ is contrasted between the first-order transition at $T = 0.9$ [panel (b)] and the second-order transition at $T = 1.0$ [panel (d)]. (c) Determination of the transition point $\Delta^{\ast}_\infty$ from the extrapolation of the finite-size pseudo-transition points $\Delta^{\ast}_L$ along the $1/L^2$ scaling line.
    } 
    \label{fig:field-mixing-example}
\end{figure}

At a first-order transition point, the probability distribution $P(\mathcal{Q})$ conjugate to the field $\lambda$ in the direction across the phase boundary is double-peaked and symmetric. On the other hand, the distribution of the operator $\mathcal{E}$ conjugate to the field $g$ along the tangent of the phase boundary is bell-shaped. Although both follow the scaling Ansatz near the tricritical point, it was demonstrated that the symmetric double-peaked structure of $P(\mathcal{Q})$ works as a sensitive indicator of the phase coexistence when computed using the Wang-Landau joint density of states~\cite{kwak2015}.

We determine the parameter set of $(s_L, \Delta^\ast_L)$ for a given temperature $T$ that produces the double-peaked symmetric shape of the distribution $P(\mathcal{Q})$. Note that the mixing parameter $s_L$ and the pseudo-transition point $\Delta^\ast_L$ exhibit system-size dependence. The other mixing parameter, $r$, does not influence the shape of $P(\mathcal{Q})$ and thus remains undetermined. For a given $T$, we solve two coupled equations with the distribution of $\mathcal{Q}$. Since the shape of the distribution is independent of both the scale and shift, we consider a variable $\tilde{\mathcal{Q}} \equiv (\mathcal{Q} - \langle \mathcal{Q} \rangle) / \sigma_{\mathcal{Q}}$ normalized with the average $\langle \mathcal{Q} \rangle$
and the standard deviation $\sigma_\mathcal{Q}$ in the numerical procedures. The first equation is the equal-population condition, 
$\sum_{\tilde{\mathcal{Q}} > 0} P(\tilde{\mathcal{Q}}) = 1/2$, which can be solved directly with discrete spectrum. The second equation is the equal-height condition, $\max_{\tilde{\mathcal{Q}} > 0} P(\mathcal{Q}) = \max_{\mathcal{Q} < 0} P(\mathcal{Q})$, where we introduce a small Gaussian broadening of $10^{-2}$ for $L\ge 16$ ($2\times 10^{-2}$ for $10 \le L \le 15$ and $3\times 10^{-2}$ for $L<10$) to evaluate the peak height. The procedure begins by setting an initial set of ($s$, $\Delta$) values in the graphical search for the parameters at which two peaks appear in $P(\tilde{\mathcal{Q}})$ and then proceeds to our numerical solver composed of nested loops. The outer loop solves the equation of the equal-height condition for the mixing parameter $s$, and the inner loop finds $\Delta$ that satisfies the equal-height condition for every value of $s$ visited in the outer loop. 

Figure~\ref{fig:field-mixing-example} exemplifies the shape of the resulting $P(\tilde{\mathcal{Q}})$ along the phase-coexistence curve and the determination of the transition point. At a low temperature deep in the first-order transition area, two sharp peaks are prominent, even for very small system sizes. The peaks broaden as the temperature increases towards the tricritical point. The pseudo-transition points $\Delta^{\ast}_L$ detected by the double-peaked structure show a clear $1/L^2$ scaling behavior expected in first-order transitions~\cite{binder84,binder87}, allowing us to determine $\Delta^{\ast}_\infty$ via the finite-size scaling formula
\begin{equation}
\label{eq:fit_delta_star}
\Delta^{\ast}_L = \Delta^{\ast}_\infty +b_{\Delta} L^{-2}.
\end{equation}
In Eq.~\eqref{eq:fit_delta_star} above, $\Delta^{\ast}_\infty$ is the desired transition field and $b_{\Delta}$ a non-universal fitting parameter. The values of $\Delta^{\ast}_\infty$ are listed in Table~\ref{tab:delta_c} for the temperatures in the first-order transition area and the vicinity of the tricritical point. 

Although the double-peaked structure is evident at temperatures both below and above the tricritical temperature $T_{\rm t}$, as illustrated for small systems in Fig.~\ref{fig:field-mixing-example}, the definite first- and second-order character of the transition can be undoubtedly distinguished by the finite-size behavior of $P(\tilde{\mathcal{Q}})$. At the first-order transition regime the peaks become sharper as $L$ increases (see also Figs.~\ref{fig:distributions}(a) and \ref{fig:distributions}(b) below). In contrast, at the second-order transition, as shown in Fig.~\ref{fig:field-mixing-example}(d), the double-peaked structure becomes less pronounced and the peak height appears to decrease with increasing system size. Within the scope of the present calculations, we are unable to make a definitive prediction regarding whether the double-peaked structure persists in the thermodynamic limit. Addressing this question would require numerical data for very large systems, which are beyond the reach of our current computational resources.

\begin{figure}
    \includegraphics[width=1.0\linewidth]{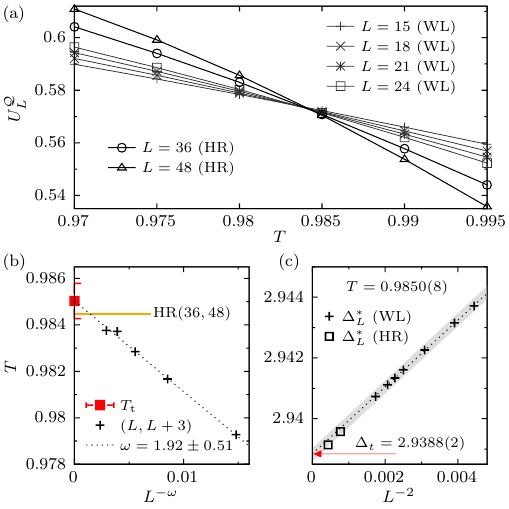}
    \caption{Location of the tricritical point: (a) The fourth-order cumulant $U^\mathcal{Q}_L$ is plotted at $\Delta^{\ast}_L(T)$. The data of $L = 36$ and $48$ are the histogram reweighting (HR) estimates. (b) The tricritical temperature $T_\mathrm{t} = 0.9850(8)$ is determined by the power-law extrapolation of the crossing points between the Wang-Landau (WL) curves of $L$ and $L + 3$. The horizontal bar indicates the crossing point of the histogram reweighting curves. (c) The tricritical field $\Delta_\mathrm{t} = 2.9388(2)$ is computed by the $1/L^2$ extrapolation of the Wang-Landau data. The markers refer to $T=0.985$ and the shaded area is bounded by the lines of $\Delta^{\ast}_L(T)$ for $T = 0.9850 \pm 0.0008$.}
    \label{fig:tricritical-point}
\end{figure}

\subsection{Locating the tricritical point}
\label{sec:TCP}

At the tricritical point, the probability distribution $P(\tilde{\mathcal{Q})}$ becomes scale-invariant. This can be assessed by probing the fourth-order cumulant
\begin{equation}
    U^\mathcal{Q}_L = 1 - \frac{\langle \tilde{Q}^4 \rangle}{3\langle \tilde{Q}^2 \rangle^2},
\end{equation}
along the phase boundary $\Delta^\ast_L(T)$ and the mixing parameter $s_L(T)$ determined in the field-mixing analysis described in Sec.~\ref{sec:field-mixing}. Figure~\ref{fig:tricritical-point}(a) illustrates the curves of $U^\mathcal{Q}_L$ as a function of the temperature $T$ computed for different system sizes. From the power-law extrapolation of their crossing points, we estimate the tricritical temperature $T_{\rm t} = 0.9850(8)$ at which the tricritical field is determined to be $\Delta_{\rm t} = 2.9488(2)$. Exactly at the tricritical point we find $U^\mathcal{Q}_L = 0.571(3)$, which is in good agreement with the estimate $0.574(2)$ reported by Wilding and Nielaba~\cite{wilding96} for the square-lattice Blume-Capel model. This finding is a clear manifestation of Ising tricritical universality. 

\begin{figure}
    \includegraphics[width=1.0\linewidth]{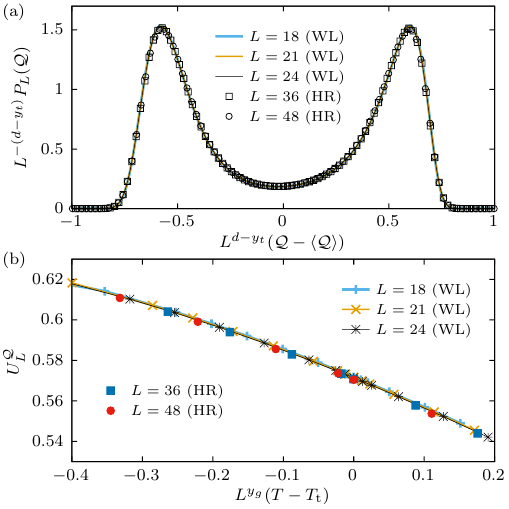}
    \caption{Test of the conjectured tricritical eigenvalue exponents. (a) Scaling collapse of the distribution $P_L(\mathcal{Q})$ with $y_t = 9/5$ at $T = 0.985$. (b) Finite-size scaling analysis of the fourth-order cumulant $U^\mathcal{Q}_L$ with $y_g = 4/5$ on the phase coexistence curve. Results obtained via Wang-Landau (WL) and histogram reweighting (HR) methods.}
    \label{fig:tricritical-exponents}
\end{figure}

As shown in Fig.~\ref{fig:tricritical-point}(b), the tricritical temperature is estimated from the value of $T_\infty$, derived from curve fitting the crossing points of adjacent curves for 
$L$ and $L+3$, where $L\in\{9,12,15,18,21\}$. The power-law fit, described by the empirical law $T_L = T_\infty + a L^{-\omega}$, aligns well with the data for smaller system sizes. However, slight deviations are observed in the data points for $L = 18$ and $21$, which is possibly due to the reduced flatness criterion we employed to address computational time constraints in the Wang-Landau calculations of the joint density of states.

We further refine our estimate of the tricritical point $(\Delta_{\rm t}, T_{\rm t})$ by conducting histogram reweighting calculations for larger system sizes of $L = 36$ and $48$. 
We prepared the histogram $H(E_J, E_\Delta; \beta_0, \mu_0)$ from Monte Carlo measurements using the Metropolis algorithm at a predetermined field $\Delta^{(0)}_L \equiv \mu_0/\beta_0$ for a given temperature $T \equiv 1/\beta_0$. To ensure accuracy, it is crucial to position $\Delta^{(0)}_L$ as close as possible to the pseudo-transition point that we aim to identify. Therefore, we set $\Delta^{(0)}_L$ to the value of $\Delta^{\ast}_{L}$, extrapolated for a larger $L$ using the fit of Eq.~\eqref{eq:fit_delta_star}, applied to the Wang-Landau data from small sizes. Thus, reweighting with $|\mu^\prime| < 0.005$ proved sufficient for our field-mixing analysis, accurately providing the mixing parameter $s$ and the pseudo-transition point for $L=36$ and $48$.

The resulting fourth-order cumulant curves, $U^\mathcal{Q}_L$, obtained using the histogram-reweighted probability distributions are shown in Fig.~\ref{fig:tricritical-point}(a). As shown in Figs.~\ref{fig:tricritical-point}(b) and (c), the crossing point between the $L = 36$ and $48$ curves falls within the error bar of our estimate for $T_{\rm t} = 0.9850(8)$. Additionally, the pseudo-transition points derived from the reweighted distributions align well with the extrapolation line of the Wang-Landau data points, supporting our estimate of $\Delta_{\rm t} = 2.9388(2)$.

\subsection{Tricritical exponents}
\label{sec:triexp}

The identification of the tricritical point in our field-mixing analysis prompts us to verify the proposed tricritical Ising exponents. Near the tricritical point, the probability distribution can be expressed as 
\begin{widetext}
\begin{equation}
\label{eq:fss_tricrit_dist}
    P_L \propto p^{\ast}_L(a_t^{-1}L^{d-y_t}\mathcal{Q},
    a_g^{-1} L^{d-y_g}\mathcal{E}, a_h^{-1} L^{d-y_h}m,  
    a_t L^{y_t} \lambda, a_g L^{y_g} g , a_h L^{y_h} h), 
\end{equation}
\end{widetext}
where $p^{\ast}_L$ is a universal scaling function, $m$ and $h$ denote the magnetization and magnetic field, respectively, and $a$'s are non-universal factors (for further details on the field-mixing method we refer the interested reader to Refs.~\cite{wilding96,bruce92,wilding92,plascak13}). 
The relevant eigenvalue exponents for the tricritical Ising universality class in two dimensions are known to be $y_t = 9/5$, $y_g = 4/5$, and $y_h = 77/40$~\cite{dennijs79,nienhuis79,pearson80,nienhuis82}.

\begin{figure}
    \includegraphics[width=1.0\linewidth]{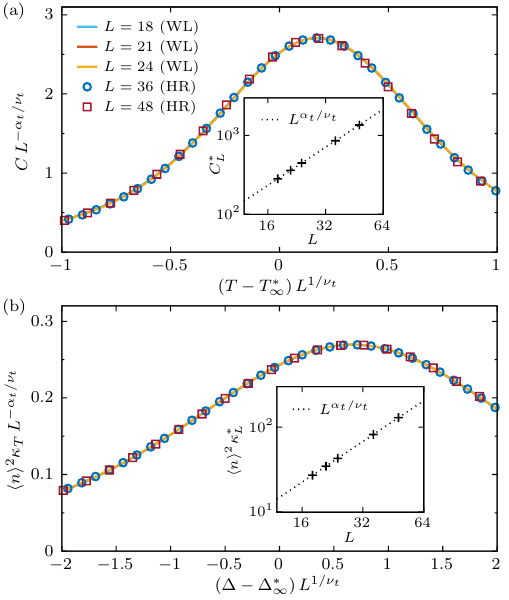}
    \caption{Finite-size scaling analysis for the tricritical exponent $\alpha_t$. (a) Specific heat $C$ as a function of $T$ with $\mu = \Delta/T$ being fixed at the Wang-Landau estimate of the tricritical value $\Delta_{\rm t}/T_{\rm t}$. The inset shows the estimate of $\alpha_t/\nu_t = 1.6059(8)$ from the scaling of the peak point $C^{\ast}_L$. (b) Isothermal compressibility $\kappa_T$ at the Wang-Landau estimate of $T_{\rm t}$. The estimate of $\alpha_t / \nu_t = 1.5950(2)$ is obtained from the scaling of the maxima. The finite-size scaling collapse is demonstrated for $T^\ast_\infty = 0.98493$, $\Delta^\ast_\infty = 2.93847$, and $\nu_t = 1/y_t$.}
    \label{fig:triexp-alpha}
\end{figure}

\begin{figure}[!ht]
    \centering
    \includegraphics[width=1.0\linewidth]{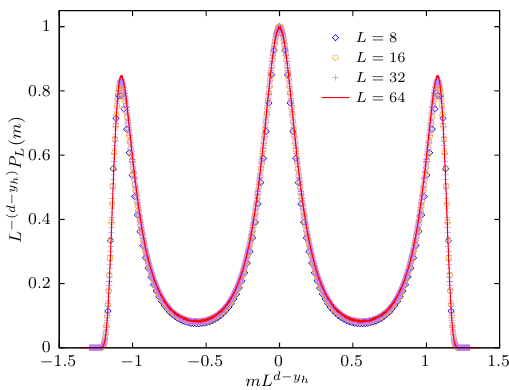}
    \caption{Finite-size scaling analysis of the magnetization distribution at the tricritical point. The temperature is fixed at $T = 0.985$. The probability distributions are generated using the Metropolis algorithm at the pseudo-transition points $\Delta = \Delta^*_L$ corresponding to the peak of the magnetic susceptibility for the examined system sizes.} 
    \label{fig:tricrit_mag_dist}
\end{figure}

At the tricritical point, the finite-size scaling Ansatz for the variable $\mathcal{Q}$ can be expressed more explicitly as 
\begin{equation}
\label{eq:fss_qdist}
    P_L(\mathcal{Q}) = L^{d-y_t} \mathcal{P}_o[L^{d-y_t}(\mathcal{Q} - \langle\mathcal{Q}\rangle)],
\end{equation}
where $\mathcal{P}_o$ is scale-invariant. To examine this scaling form, it is essential to have both mixing parameters, $r$ and $s$, to compute $\mathcal{Q}$. While we have obtained $s$ from the field-mixing analysis, the other parameter $r$ can be determined from the phase coexistence curve in the $\beta-\mu$ plane. In Fig.~\ref{fig:phase-diagram}(c) we obtained $r = -3.2009(3)$ from the slope of the curve at the tricritical point. Figure~\ref{fig:tricritical-exponents}(a) shows an excellent finite-size scaling collapse of $P_L(\mathcal{Q})$ with the given exponent $y_t = 9/5$ at $T = 0.985$, verifying the conjectured tricritical exponent in the triangular Blume-Capel ferromagnet.     

The finite-size scaling expression~\eqref{eq:fss_tricrit_dist} also suggests that the fourth-order cumulant $U^\mathcal{Q}_L$ would follow the form $U^\mathcal{Q}_L = u_o(L^{y_g}g)$ if computed along the phase boundary of $\lambda = 0$. The scaling field $g$ is the deviation from the tricritical point in the direction tangent to the coexistence curve. As we have shown in the inset of Fig.~\ref{fig:phase-diagram}, the curve is almost linear near the tricritical point in the $\beta$-$\mu$ plane. Thus, we can express $g \propto (T - T_{\rm t})$ in its leading-order approximation for $T \approx T_{\rm t}$, which enables us to reformulate the finite-size scaling form as follows 
\begin{equation}
\label{eq:fss_uq}
    U^\mathcal{Q}_L = \mathcal{U}_o[L^{y_g}(T-T_{\rm t})].
\end{equation}
This specific scaling form is illustrated in Fig.~\ref{fig:tricritical-exponents}(b), where data points from different system sizes fall nicely onto a common curve with the exponent $y_g = 4/5$.

In addition, we consider the specific heat and isothermal compressibility to measure the tricritical exponent $\alpha_t$~\cite{kwak2015}. The finite-size scaling analysis can be done along each axis of the $\beta-\mu$ plane near the tricritical point. Along the temperature axis, the specific heat at fixed $\mu$ is defined by the fluctuations of $\epsilon = L^{-d}E_J$ as 
\begin{equation}
    C = \frac{L^d}{T^2} \left( \langle \epsilon^2 \rangle - \langle \epsilon \rangle^2 \right),
\end{equation}
which is examined for a range of temperatures $T$ by setting $\Delta = T \mu$. We have $\mu = \Delta / T$ fixed at the Wang-Landau estimate of the tricritical value $\Delta_\mathrm{t}/T_\mathrm{t}$. Note that for the larger system sizes $L = 36$ and $48$, the multiple histogram method was used in order to explore the two-dimensional space of $(\Delta, T)$ by assuming the Poissonian variation in the histogram~\cite{barkema_book}. On the other hand, the isothermal compressibility of the non-zero spin density $n$,
\begin{equation}
    \kappa_T = \frac{L^d}{T} \frac{\langle n^2 \rangle - \langle n \rangle^2}{\langle n \rangle^2},
\end{equation}
is evaluated at a fixed $T = T_{\rm t}$ for varying $\Delta$.
In the tricritical area both quantities are expected to exhibit a similar finite-size scaling behavior, namely 
\begin{equation}
    C = L^{\alpha_t / \nu_t} \mathcal{C}^o[ (T-T_{\rm t}) L^{1/\nu_t}],
    \end{equation}
and
\begin{equation}
    \langle n \rangle^2 \kappa_T = L^{\alpha_t / \nu_t} \mathcal{K}^o[(\Delta-\Delta_{\rm t}) L^{1/\nu_t}],
\end{equation}
respectively, where $\mathcal{C}^o$ and $\mathcal{K}^o$ are universal functions. The exponents $\alpha_t$ and $\nu_t$ are related to the tricritical eigenvalue exponent $y_t$ as $\alpha_t / \nu_t = -d + 2y_t$, and the hyperscaling relation $d \nu_t = 2 - \alpha_t$ indicates $\nu_t = 1/y_t$. Thus, the conjectured value of $\alpha_t / \nu_t$ is $8/5$ in the tricritical Ising universality class. 

Figure~\ref{fig:triexp-alpha} displays the finite-size scaling analysis for $\alpha_t / \nu_t$ based on the specific heat (panel (a)) and the isothermal compressibility of non-zero spin density (panel (b)). Examining the power-law behavior of the maxima of $C$ and $\langle n \rangle^2 \kappa_T$, we measured $\alpha_t / \nu_t = 1.6058(8)$ and $1.5950(2)$, respectively, both in excellent agreement with the conjectured value $8/5$. Also, the location of the peak showed a clear scaling behavior of the form $x_L = x_\infty + a L^{-y_t}$, where $x_L \equiv T^\ast_L$ and $\Delta^\ast_L$ for the maxima of specific heat and compressibility, respectively. From the linear fit and using $y_t = 9/5$ we obtained the pseudo-transition point $\Delta^\ast_\infty = 2.39847(1)$ and $T^\ast_\infty = 0.98483(1)$, which agrees very well with the Wang-Landau estimate of the tricritical point from the field-mixing analysis, i.e., $\Delta_\mathrm{t} = 2.9388(2)$ and $T_\mathrm{t} = 0.9850(8)$. Using the measured values of $\alpha_t / \nu_t$ and $x_\infty$, our finite-size scaling tests across various system sizes demonstrate a perfect collapse onto a single curve in each scaling axis for the specific heat and isothermal compressibility.

Finally, let us discuss the tricritical signatures of the magnetization distribution. A hallmark of the tricriticality in the magnetization distribution is the emergence of three peaks~\cite{wilding96}. Moreover, the corresponding probability density function of magnetization per spin, $m \equiv L^{-d}\sum_x \sigma_x$, has a universal finite-size scaling form,  
\begin{equation}
    P_{L}(m) = L^{d-y_h} \mathcal{P}_o(L^{d-y_h}m),
\end{equation}
where the exponent $y_h = 77/40$ is conjectured for the tricritical Ising universality class. In the seminal work of Wilding and Nielaba~\cite{wilding96}, it was discussed mainly in the model of spin fluids, while it was compared with the square-lattice Blume-Capel model of a particular system size of $L = 40$. Here, our combined numerical efforts in the triangular Blume-Capel model confirm these signatures with comprehensive finite-size scaling analysis. Figure~\ref{fig:tricrit_mag_dist} identifies the emergence of the three peaks in the magnetization distribution at the tricritical temperature, and along the pseudo-transition point of $\Delta$ for finite $L$, the triple-peaked distribution exhibits an excellent finite-size scaling collapse of different system sizes falling onto a single curve with the conjectured exponent $y_h = 77/40$. Therefore, we have successfully verified all three conjectured eigenvalue exponents, $y_t$, $y_g$, and $y_h$, at our estimate of the tricritical point.

\begin{figure}
    \includegraphics[width=1.0\linewidth]{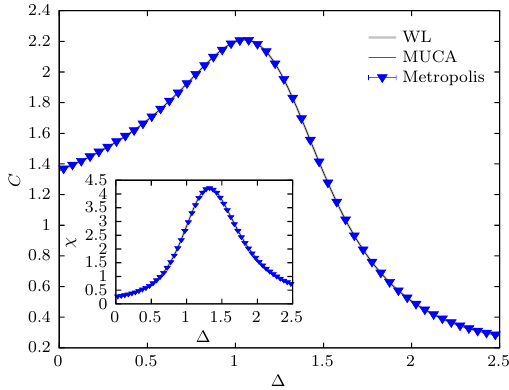}
    \caption{Specific heat (main panel) and magnetic susceptibility (inset) curves as a function of $\Delta$ for the $L = 16$ triangular Blume-Capel model at $T = 2.395$ (in the second-order transition regime). Different sets of numerical data obtained via Wang-Landau (WL), multicanonical (MUCA), and Metropolis simulations are compared.} 
    \label{fig:comparison}
\end{figure}

\section{Transition features and finite-size scaling}
\label{sec:transitions}

In the following, we discuss the transition characters of the spin-$1$ triangular Blume-Capel model by conducting large-scale multicanonical simulations and a customized finite-size scaling analysis along both sides of the tricritical point on the $(\Delta, T)$ phase boundary.

\subsection{Multicanonical simulations}
\label{sec:MUCA}

The multicanonical (MUCA) method~\cite{berg92} consists of a substitution of the Boltzmann factor $e^{-\beta E}$ with weights that are iteratively modified to produce a flat histogram, usually in energy space. This ensures that suppressed states, such as those in the co-existence region in an (effectively) first-order transition scenario, can be reliably sampled, and a continuous reweighting to arbitrary values of the external control parameter becomes possible~\cite{janke03,gross18}. Due to the two-parametric nature of the density of states $\Gamma(E_J, E_\Delta)$ in the Blume-Capel model, the process was applied only to the crystal-field part $E_\Delta$ of the energy. This allowed us to reweight to arbitrary values of $\Delta$ while keeping the temperature fixed. 
Starting from the partition function of Eq.~\eqref{eq:Z} we can write
\begin{equation}
    {\cal Z}_\mathrm{MUCA}=\sum_{E_J, E_\Delta} \Gamma(E_J, E_\Delta) e^{-\beta E_J} W\left(E_\Delta\right),
\label{eq:zng}
\end{equation}
where the Boltzmann weight associated with the crystal-field part of the energy has been generalized to $W\left(E_\Delta\right)$. For a flat marginal distribution in $E_\Delta$, it should hold that
\begin{equation}
    W(E_{\Delta}) \propto {\cal Z}_\mathrm{MUCA} \left[ \sum_{E_J} \Gamma(E_J, E_\Delta)  e^{-\beta E_J}  \right]^{-1}.
\label{eq:wmuca}
\end{equation}
In order to iteratively approximate the generalized weights $W\left(E_\Delta\right)$, we sampled histograms of the crystal-field energy. Supposing that at the $n^\text{th}$ iteration, a histogram $H^{(n)}(E_\Delta)$ was sampled, then its average should depend on the weight of the iteration $W^{(n)}\left(E_\Delta\right)$ as
\begin{equation}
    \langle H^{(n)}(E_\Delta)\rangle \propto \sum_{E_J} \Gamma(E_J, E_\Delta) e^{-\beta E_{J}} W^{(n)}(E_\Delta).
\label{eq:hmuca}
\end{equation}
From Eqs.~\eqref{eq:wmuca} and \eqref{eq:hmuca}, it follows that $\langle H^{(n)}(E_\Delta)\rangle\propto W^{(n)}(E_\Delta) / W(E_\Delta)$. This justifies our weight modification scheme $W^{(n+1)}\left(E_\Delta\right) = W^{(n)}\left(E_\Delta\right)/H^{(n)}(E_\Delta)$ for iterations to approximate $W(E_\Delta)$ producing a flat histogram.
The iterations stop when the histogram satisfies a suitable flatness criterion. We tested the flatness of the histogram using the Kullback-Leibler divergence~\cite{kullback51, gross18}, which is adequate for our distributions that do not have a long tail of near-zero values. The weights are fixed after the iterations and then used for the production runs. 

As has been shown in detail in Refs.~\cite{gross18,zierenberg13}, the multicanonical method can be adapted for the use on parallel machines by performing the sampling of histograms in parallel, with each parallel worker using the same weights but a different (independent) pseudo-random number sequence. The accumulated histogram can then be used to update the weights, keeping communication between the parallel parts of the code minimal. This scheme has been successfully applied for the study of spin systems in the past, including the spin-$1$ Blume-Capel and Baxter-Wu models~\cite{zierenberg2015,zierenberg17,fytas18,fytas22,vasilopoulos22,macedo23}. Here we performed our simulations on an Nvidia TURING RTX2080 or on a PASCAL GTX1080T, depending on which was available, giving us access to simulating $47\,104$ and $57\,344$ independent copies of the system. 
At each time, a subset of these threads are actually running in parallel, while the excess in the number of parallel tasks is employed to improve the latencies due to memory accesses~\cite{weigel18}. 

Since the multicanonical method allows unrestricted reweighting to any value of $\Delta$, the canonical expectation value of an observable $O = O(\{\sigma\})$ at a fixed temperature can be computed for an arbitrary $\Delta$ as
\begin{equation}
	\langle O\rangle_\Delta
	=
	\frac{ \left\langle O(\{\sigma\})\,e^{-\beta\Delta E_{\Delta}(\{\sigma\})} W^{-1}\left(E_{\Delta}\right) \right\rangle_\mathrm{MUCA}}
	{ \left\langle e^{-\beta\Delta E_{\Delta}(\{\sigma\})} W^{-1}\left(E_{\Delta}\right) \right\rangle_\mathrm{MUCA} }.
	\label{eq:muca-reweight}
\end{equation}
Evaluating Eq.~\eqref{eq:muca-reweight} requires high-precision summation of the exponentials to prevent numerical overflow and round-off errors. We addressed these challenges using the {\tt GNU MPFR} library~\cite{MPFR} with allocating a sufficient number of bits for numerical representation. To identify maxima along the reweighted curves, we implemented the Brent method from the GNU Scientific Library.

In our multicanonical simulations, we considered triangular lattices with periodic boundary conditions and linear sizes in the range $12 \leq L \leq 128$. To thoroughly probe the characteristics of the transition on both sides of the phase boundary across the tricritical point, we performed simulations at various temperatures from $T = 2.395$ in the second-order transition regime down to $T = 0.75$ in the first-order regime. 
We note here that the multicanonical simulations become prohibitively demanding in the low-temperature regime, a fact that is also reflected in the statistical fluctuations of the large-$L$ data illustrated in Figs.~\ref{fig:tension} and \ref{fig:shift_eqh} for $T = 0.75$. Last but not least, for all fits performed throughout this paper, we restricted ourselves to data with $L\geq L_{\rm min}$, adopting the standard $\chi^{2}$ test for goodness of the fit. Specifically, we considered a fit as being acceptable only if $10\% < Q < 90\%$, where $Q$ is the quality-of-fit parameter~\cite{press92}.

\begin{figure}[!ht]
    \centering
    \includegraphics[width=1.0\linewidth]{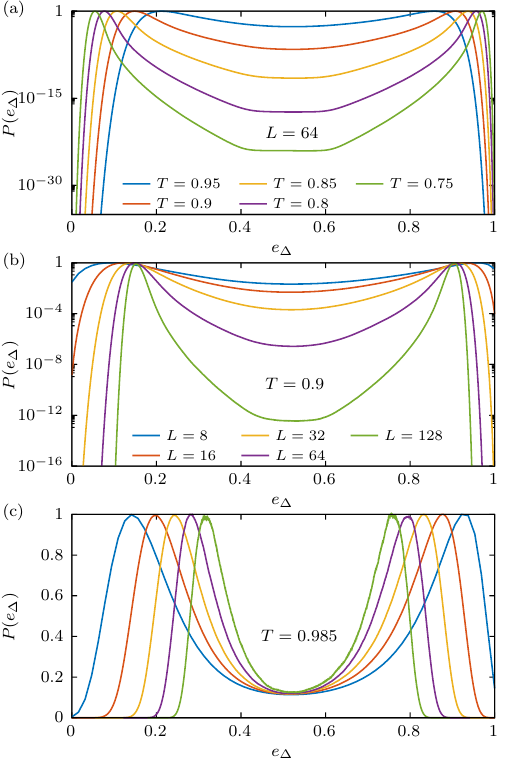}
    \caption{Probability density functions, $P(e_{\Delta})$, obtained from multicanonical simulations at: (a) Selected temperatures within the first-order transition regime for a system with linear size $L = 64$, (b) the characteristic temperature $T = 0.9$ and a wide range of system sizes, and (c) the estimated tricritical temperature ($T = T_{\rm t} = 0.985$) for the same set of system sizes as in panel (b). Note the logarithmic scale of the vertical axis in panels (a) and (b).} 
    \label{fig:distributions}
\end{figure}

\subsection{Observables}
\label{sec:obs}

Using the multicanonical simulations, we measured the statistics of various observables, including the energy $E = E_J + \Delta E_\Delta$ and the order parameter $m$, leading us to the estimate of the magnetic susceptibility 
\begin{equation}\label{eq:susceptibility}
 \chi = \beta N\left[\langle m^{2}\rangle - \langle m\rangle^{2}\right],
\end{equation}
and the specific heat
\begin{equation}
	C = \beta^2 N \left[ \left\langle E^2 \right\rangle - \left\langle E \right\rangle^2 \right].
\end{equation}
Additionally, for the direct estimate of the critical exponent $\nu$ from the finite-size scaling behavior, we computed the logarithmic derivative of $n^{\rm th}$ power of the order parameter~\cite{ferrenberg91,caparica00,malakis09}
\begin{equation}
\label{eq:log_der}
	K^{(n)} \equiv \frac{\partial\ln{\langle m^n \rangle}}{\partial T}
	= \left[\frac{\left\langle m^n E \right\rangle}{\langle m^n \rangle} - \left\langle E \right\rangle  \right].
\end{equation}

\begin{figure}[!ht]
    \centering
    \includegraphics[width=1.0\linewidth]{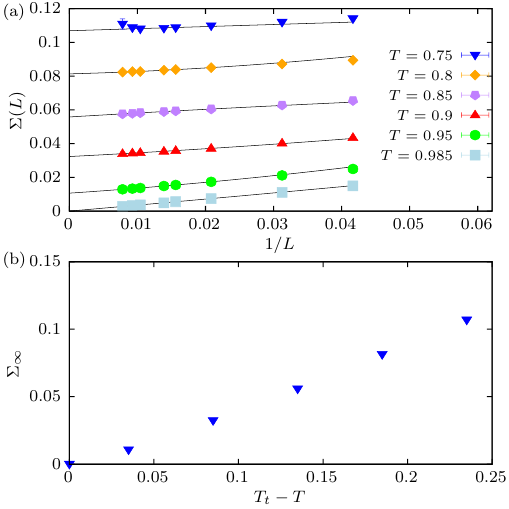}
    \caption{(a) Limiting behavior of the surface
     tension $\Sigma(L)$, specifically at the tricritical point ($T = T_{\rm t} = 0.985$) and well inside the first-order transition regime for a series of selected temperatures, as indicated. Data are obtained from the equal height $P(e_{\Delta})$ distributions of the multicanonical simulations and results for $L > 16$ are shown. The solid lines are fits of the form~\eqref{eq:tension}, including a second-order correction term and $L_{\rm min} = 48$. (b) Temperature dependence of $\Sigma_{\infty}$ as obtained from the fits documented in panel (a), measured in relation to the distance from the tricritical point, $T_{\rm t} - T$.} 
    \label{fig:tension}
\end{figure}

At this point, having defined some of the main observables of interest, we call attention to a comparative check that highlights the numerical accuracy of the simulation methods implemented in this work. Figure~\ref{fig:comparison} contrasts the specific heat and magnetic susceptibility curves of a $L = 16$ triangular Blume-Capel system at the temperature $T = 2.395$ obtained via three different protocols: (i) the Wang-Landau joint density of states outlined in Sec.~\ref{sec:WL}, (ii) the multicanonical scheme sketched here, and (iii) the well-known Metropolis algorithm~\cite{barkema_book}. Without doubt, Fig.~\ref{fig:comparison} showcases a perfect matching among the three datasets.

As we are mostly interested in the first-order characteristics of the transition, following the prescription of Ref.~\cite{vasilopoulos22}, we constructed the probability density function of $E_\Delta$ per spin, $P(e_{\Delta})$, where $e_{\Delta} = E_{\Delta}/L^2$. It is well known that a double-peaked structure in the probability density function in finite systems is a precursor of the two delta functions expected in the thermodynamic limit for a first-order phase transition~\cite{binder84,binder87}. Typical (normalized to unity) probability distributions $P(e_{\Delta})$ for various system sizes and temperatures in the first-order transition regime and at the proposed tricritical point are shown in Fig.~\ref{fig:distributions}. 
\begin{figure}[!ht]
    \centering
    \includegraphics[width=1.0\linewidth]{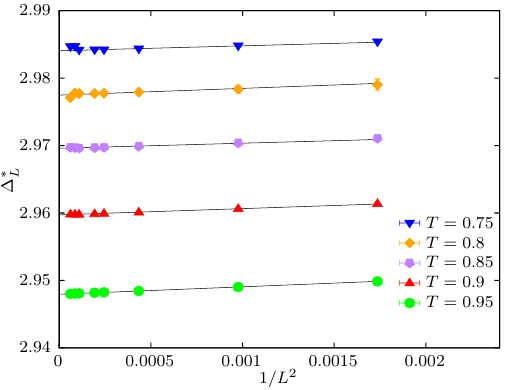}
    \caption{Shift behavior of pseudo-transition fields $\Delta_{L}^{\ast}$ determined by 
    the equal-height peaks of the probability density function $P(E_{\Delta})$ for the range of temperatures considered in Fig.~\ref{fig:distributions}. Results for $L > 16$ are shown. The solid lines are fits of the form~\eqref{eq:fit_delta_star} to the data points of $L \ge L_\mathrm{min} = 24$. The extrapolated values of $\Delta_{\infty}^{\ast}$ are listed in Table~\ref{tab:delta_c}.} 
    \label{fig:shift_eqh}
\end{figure}
A few comments are in order: (i) In the first-order transition regime, the barrier increases upon lowering the temperature, reaching the order of $\sim 10^{-24}$ for $T = 0.75$ [panel (a)]. (ii) Note the strong enhancement of the barrier with increasing system size [panel (b)], exhibiting a major suppression of intermediate states, which is common for first-order phase transitions. (iii) At the proposed tricritical point, [panel (c)], the distributions still feature a double-peaked structure, yet with a much smaller barrier which seems to be independent of system size. A more illuminating discussion on the asymptotic behavior of these observables regarding the distributions of Fig.~\ref{fig:distributions} is given below.

\subsection{Finite-size scaling analysis}
\label{sec:FSS}

\subsubsection{First-order transition regime}
\label{subsec:first-order}

The observations in Fig.~\ref{fig:distributions} naturally lead us to investigate the systematic behavior of the surface tension that may characterize the transition as suggested by Lee and Kosterlitz~\cite{lee90,lee91}. The multicanonical method is instrumental for this purpose since it allows the direct estimation of the barrier associated with suppressing intermediate states during a first-order phase transition. 
Considering distributions with two peaks of equal height (eqh)~\cite{borgs92}, such as the ones shown in Fig.~\ref{fig:distributions}, allows one to extract the free-energy like barrier in the $e_{\Delta}$-space,
\begin{equation}
	\Delta F(L) = \frac{1}{2\beta\Delta}\ln{\left(\frac{P_{\rm
			max}}{P_{\rm min}}\right)_{\rm eqh}},
\end{equation}
where $P_{\rm max}$ and $P_{\rm min}$ are the maximum and local minimum of the distribution $P(e_{\Delta})$, respectively. 
\begin{figure}[!ht]
    \centering
    \includegraphics[width=1.0\linewidth]{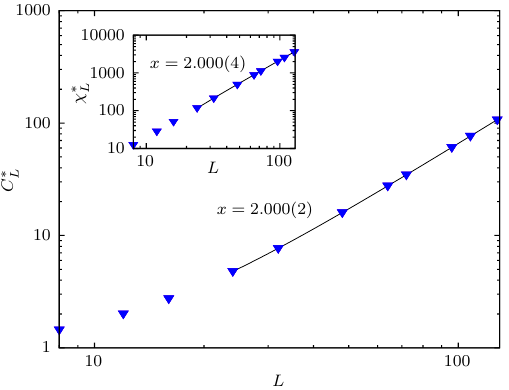}
    \caption{Finite-size scaling behavior of the specific heat (main panel) and magnetic susceptibility (inset) maxima, located along the $\Delta$ axis at the fixed temperature $T=0.9$ in the first-order transition regime. The solid lines indicate fits of the forms~\eqref{eq:cscaling} and \eqref{eq:chiscaling} respectively, for
    $L \ge L_\mathrm{min}= 24$.
    } 
    \label{fig:fss_cx_1st}
\end{figure}
The resulting barrier connects a spin-$0$ dominated regime ($e_{\Delta}$ small) and a spin-$\pm1$ rich phase ($e_{\Delta}$ large). The corresponding surface tension $\Sigma(L) = \Delta F(L)/L$ is expected to scale as
\begin{equation}
\label{eq:tension}
\Sigma(L)  = \Sigma_{\infty} + c_1L^{-1} + \mathcal{O}\left(L^{-2}\right), 
\end{equation}
in two dimensions, possibly with higher-order corrections~\cite{Nussbaumer2006,Nussbaumer2008,Bittner2009}. We point out here the recent numerical evidence of Ref.~\cite{mendes24} that relate the surface tension $\Sigma$ to an
interface between a spin-$\pm1$ phase separated by a surfactant layer of zeros, leading to a
reduced effective surface tension~\cite{schick86,cirillo96}. The scaling behavior of the surface tension is depicted in Fig.~\ref{fig:tension}(a) for the complete set of temperatures considered during the multicanonical simulations and the solid lines are fits of the form~\eqref{eq:tension}. The retrieved $\Sigma_{\infty}$ values are plotted against the temperature difference from the tricritical point ($T_{\rm t} - T$) in Fig.~\ref{fig:tension}(b), featuring an almost linear increase in $\Sigma_{\infty}$ as the temperature decreases towards the limit $T\rightarrow 0$. 
This result agrees nicely with the observations by Jung and Kim for the square-lattice Blume-Capel model; see Fig.~4(a) in Ref.~\cite{kim17}. Additionally, we verify that at the proposed tricritical temperature $T = 0.985$, the surface tension vanishes to a limiting value of $\Sigma_{\infty} = 3\times 10^{-6} \pm 2\times 10^{-5}$.

We now examine the system-size scaling behavior of the pseudo-transition field $\Delta_{L}^{\ast}$ obtained using the probability density function $P(e_{\Delta})$. We determine $\Delta_{L}^{\ast}$ by the value of the crystal field at which the two peaks of equal height appear in $P(e_{\Delta})$, as illustrated in Fig.~\ref{fig:distributions}. Figure~\ref{fig:shift_eqh} presents the behavior of $\Delta_{L}^{\ast}$ for selected temperatures in the first-order transition regime. 
Our data indicate an excellent fit to the expected first-order behavior of the scaling form~\eqref{eq:fit_delta_star} as demonstrated by the solid lines in Fig.~\ref{fig:shift_eqh}. The extrapolated values of the transition field $\Delta_{\infty}^{\ast}$ obtained here from $P(e_{\Delta})$ confirm to a high numerical accuracy the transition points obtained from the field-mixing analysis in Sec.~\ref{sec:field-mixing}. 
\begin{figure}[!ht]
    \centering
    \includegraphics[width=1.0\linewidth]{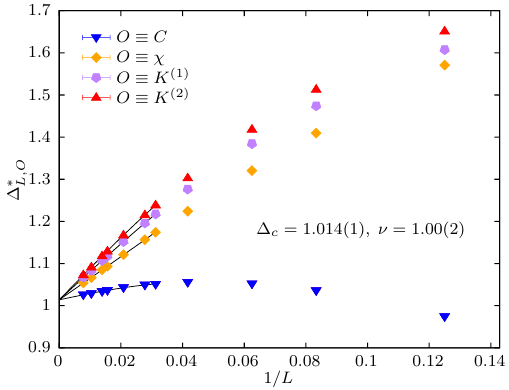}
    \caption{Shift behavior of several pseudo-critical temperatures, as defined in the main text, obtained from simulations at $T = 2.395$ in the second-order transition regime.} 
    \label{fig:shift}
\end{figure}
Table~\ref{tab:delta_c} provides the comparison between the two different methods, showing that a deviation in $\Delta_{\infty}^{\ast}$ is about $10^{-4}$ or less. 

The strong agreement in $\Delta_{\infty}^{\ast}$ between the two different approaches suggests that the parameter
$s$ in the field-mixing analysis may have a minor role in determining the transition point within the first-order regime. The probability density function $P(e_{\Delta})$ corresponds to the case of $s=0$ for the scaling operator 
$\mathcal{Q}$. Although the value of $s$ determined in the field-mixing analysis varied with system size and temperature, it was generally around $0.01$
in this model. Notably, when considering only the equal height condition, we observed a weak dependence of $\Delta^\ast_L$ on $s$. In contrast, the fourth-order cumulant $U_L^\mathcal{Q}$ becomes increasingly sensitive to $s$ as it approaches the tricritical point, highlighting the importance of the field-mixing analysis for accurately locating this point.

To conclude this section, we will focus on the temperature $T = 0.9$ to explore further aspects of the characteristic scaling associated with first-order phase transitions. In the specific heat and magnetic susceptibility, we expect that their maxima along the $\Delta$ axis behave as $C^{\ast}_L \sim L^{d}$ and $\chi^{\ast}_L \sim L^{d}$ in the leading order of $L$, respectively. Scaling corrections at first-order transitions are in inverse integer powers of the volume~\cite{janke1,janke2}, and thus one may write the corresponding scaling behavior as
\begin{equation}
    C^{\ast}_L = b_C L^{x} \left(1+b^\prime_C L^{-2}\right) \label{eq:cscaling}
\end{equation}
    and
\begin{equation}
    \chi^{\ast}_L = b_\chi L^{x} \left(1+b^\prime_\chi L^{-2}\right), \label{eq:chiscaling}
\end{equation}
where $\{b_{C}, b^\prime_C, b_{\chi}, b^\prime_\chi\}$ are non-universal amplitudes. This scaling Ansatz provides an excellent fit to our data for both observables, as shown in Fig.~\ref{fig:fss_cx_1st}. 
\begin{figure}[!ht]
    \centering
    \includegraphics[width=1.0\linewidth]{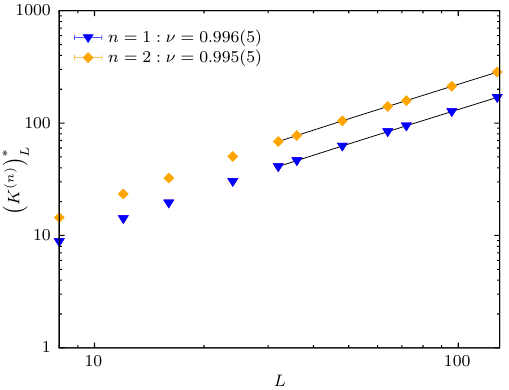}
    \caption{Finite-size scaling behavior of the maxima of the logarithmic derivatives of $n^{\rm th}$ power of the order parameter with $n=1$ and $2$ in a double logarithmic scale at $T = 2.395$ in the second-order transition regime.} 
    \label{fig:log_der}
\end{figure}
From the data fitting with $L_\mathrm{min} = 24$, we obtain  
$x = 2.000(2)$ and $2.000(4)$ for the specific heat and the magnetic susceptibility, respectively, which are in perfect agreement with the expected value $d = 2$ of the first-order transition in this model. Note that due to the value of $x\approx 2$, the $1/L^2$ correction simply becomes an additive constant. 

\subsubsection{Second-order transition regime}
\label{subsec:second-order}

As already mentioned above, the Blume-Capel model for crystal fields $\Delta < \Delta_{\rm t}$ undergoes a second-order phase transition between the ferromagnetic and paramagnetic phases that falls into the universality class of the classical Ising ferromagnet. This has been affirmed in several numerical studies of the model focusing mainly on the square lattice -- see for example Ref.~\cite{zierenberg17} and references therein -- but also on the triangular lattice~\cite{fytas_BC}. In the following we outline the finite-size scaling analysis of our high-precision numerical data obtained from multicanonical simulations at $T = 2.395$ that corroborates previous results on the model.

\begin{figure}
    \includegraphics[width=1.0\linewidth]{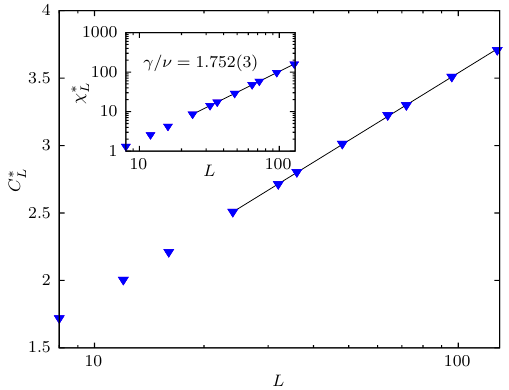}
    \caption{Finite-size scaling behavior of the maxima of the specific heat (main panel) and magnetic susceptibility (inset) at $T = 2.395$ in the second-order transition regime. Note the horizontal logarithmic scale in the main panel and the double logarithmic scale in the corresponding inset.} 
    \label{fig:fss_cx}
\end{figure}

We start in Fig.~\ref{fig:shift} with the shift behavior of several pseudo-critical temperatures corresponding to the peaks of the specific heat ($O \equiv C$), magnetic susceptibility ($O\equiv \chi$), and logarithmic derivatives of the order parameter ($O \equiv K^{(n)}$, for $n = 1$ and $2$). These scale with the system size as
\begin{equation}
\label{eq:scaling_delta_second_order} \Delta_{L,O}^{\ast}=\Delta_{\rm c}+b_{\Delta_{O}}L^{-1/\nu}(1+b^\prime_{\Delta_{O}}L^{-\omega}),
\end{equation}
with $\{b_{\Delta_{O}}, b^\prime_{\Delta_{O}}\}$ non-universal constants (similarly for the fitting coefficients $\{b^\prime_{m,(n)}, b_{C}, b^\prime_{C}, b^\prime_{\chi}\}$ used in the scaling Eqs.~\eqref{eq:K_scaling} - \eqref{eq:chi_scaling} below) and $\omega$ the universal corrections-to-scaling exponent, fixed to the Ising value $7/4$ (see, for example, the discussion in the supplementary material of Ref.~\cite{shao16} and comment~\cite{comment2}). 
Simultaneously fitting the numerical data for the larger system sizes studied ($L_{\rm min} = 32$) to the power-law~\eqref{eq:scaling_delta_second_order} we obtain $\Delta_{\rm c} = 1.014(1)$ (see also Table~\ref{tab:delta_c}) and $\nu = 1.00(2)$, the latter in very good agreement with the exact value $\nu = 1$ of the Ising ferromagnet. Figure~\ref{fig:log_der} yields a supplementary estimation of the correlation-length's exponent $\nu$ via the finite-size scaling behavior of the logarithmic derivatives of the order parameter. In particular, we demonstrate in Fig.~\ref{fig:log_der} the cases with $n=1$ and $2$ of Eq.~\eqref{eq:log_der}. Fittings of the form~\cite{ferrenberg91}
\begin{equation}
\label{eq:K_scaling} \left (K^{(n)} \right )^{\ast}_{L} = L^{1/\nu}(1+b^\prime_{m,(n)}L^{-\omega})
\end{equation}
(again with $L_{\rm min} = 32$) give an average estimate of $\nu = 0.996(5)$, in excellent agreement with the Ising value. Finally, in  Fig.~\ref{fig:fss_cx} we depict the scaling behavior of the maxima of the specific heat (main panel) and the magnetic susceptibility (inset). Following the pioneering work of Ferdinand and Fisher~\cite{fisher69} the maxima of the specific heat (close to the critical point) obey the following finite-size scaling expansion
\begin{equation}
\label{eq:c_scaling} C^{\ast}_{L} = b_{C}+b^\prime_{C}\ln{(L)},
\end{equation}
whereas the maxima of the magnetic susceptibility scale with system size according to the usual form
\begin{equation}
\label{eq:chi_scaling} \chi^{\ast}_{L} = L^{\gamma/\nu}(1+b^\prime_{\chi}L^{-\omega}).
\end{equation}
In both panels of Fig.~\ref{fig:fss_cx} the solid lines are fits of the form~\eqref{eq:c_scaling} and \eqref{eq:chi_scaling} with the cut-off of $L_{\rm min} = 24$. A clear logarithmic divergence is observed for the specific-heat data and the computation of the magnetic exponent ratio $\gamma/\nu = 1.752(3)$ is in excellent agreement with the value $7/4$ of the two-dimensional Ising universality class.

\begin{table*}[tb!]
    \begin{ruledtabular}
        \caption{A collection of transition/critical  ($\Delta_{\infty}^{\ast}/\Delta_{\rm c}$) fields for the spin-$1$ triangular Blume-Capel ferromagnet. Several results obtained in the current work from two-parametric Wang-Landau (WL) (second column) and multicanonical (MUCA) simulations (third column) at fixed temperatures (first column) are shown, including earlier estimates of critical temperatures ($T_{\rm c}$) (fourth column) from Wang-Landau simulations at fixed values of $\Delta$ (fifth column)~\cite{fytas_BC}. The last column indicates the order of the transition for the particular values of the temperature/crystal field considered during the course of the simulations. The location of the tricritical point that separates the two regimes is highlighted.}
            \label{tab:delta_c}
        \begin{tabular}{lcclcc}
            $T$ & $\Delta^{\ast}_\infty / \Delta_{\rm c}$ (WL) & $\Delta^{\ast}_\infty / \Delta_{\rm c}$ (MUCA) & $T_{\rm c}$ (WL) &  $\Delta$ & Order of transition\\
            \hline
            0.500 & 2.998691(8) &  &  & & First\\
            0.550 & 2.997482(6) &  & & & First\\ 
            0.600 & 2.995600(5) &  & & & First\\ 
            0.650 & 2.992852(4) &  & & & First\\ 
            0.700 & 2.989055(4) &  & & & First\\ 
            0.750 & 2.984020(4) & 2.98406(3)  & & & First\\
            0.800 & 2.977568(3) & 2.97758(7)  & & & First\\
            0.850 & 2.969520(2) & 2.96960(3)  & & & First\\
            0.900 & 2.959721(4) & 2.95972(1)  & & & First\\
            0.950 & 2.94808(1)  & 2.94795(3) &  & & First\\
            0.960 & 2.94553(1)  & &  & & First\\
            0.965 & 2.94422(2)  & &  & & First\\
            0.970 & 2.94290(2)  & &  & & First\\
            0.975 & 2.94157(2)  & &  & & First\\
            0.980 & 2.94021(2)  & &  & & First\\  \hline
            {\bf 0.9850(8)} & {\bf 2.9388(2)} &  & & & {\bf Tricritical}\\ \hline
            0.990 & 2.93745(3)  & &  & & Second\\
            0.995 & 2.93605(3)  & &  & & Second\\
            1.000 & 2.93464(3)  & &  & & Second\\ 
            &             &  & 1.372(4) & 2.75 & Second\\
            &             &  & 1.639(3) & 2.50 & Second\\
            &             &  & 1.978(2) & 2.00 & Second\\
             &             &  & 2.210(3)& 1.50  & Second\\
             &             &  & 2.395(2) & 1.00 & Second\\  
             2.395 &            & 1.014(1)\footnote{For comparison, note the inverse computation of $T_{\rm c} = 2.395(2)$ via Wang-Landau
            simulations at $\Delta = 1$~\cite{fytas_BC}.}& & & Second \\ 
             &             &  & 2.549(3) & 0.50 & Second\\
             &             &  & 2.675(2) & 0.00 & Second\\
        \end{tabular}
    \end{ruledtabular}
\end{table*}

\section{Summary and Outlook}
\label{sec:discussion}

In this paper, we presented a comprehensive numerical and finite-size scaling investigation of the spin-$1$ Blume-Capel ferromagnet on the triangular lattice. In the first part of our study, we mapped the detailed line of first-order phase transitions, filling in a previously underexplored region of the phase diagram at low temperatures. Using Wang-Landau simulations to measure the joint density of states, along with histogram reweighting based on the Metropolis protocol and field mixing, we achieved a high-accuracy determination of the first-order transition line -- a summary of transition points is provided in Table~\ref{tab:delta_c} -- and of the tricritical point at $(\Delta_{\rm t}, T_{\rm t}) = (2.9388(2), 0.9850(8))$. Additionally, via our combined numerical scheme we confirmed the tricritical Ising exponents $y_{t} = 9/5$, $y_{g} = 4/5$, and $y_h = 77/40$, up to a very good numerical accuracy. In the second part, we conducted extensive multicanonical simulations to investigate the finite-size scaling properties of the model in both its first- and second-order transition regimes. Our analysis yielded high-accuracy estimates of transition and critical fields, which can also be found in Table~\ref{tab:delta_c}. These align well with those stemming from the two-parametric Wang-Landau simulations carried out in the first part of our study, as well as with previous results reported in the literature~\cite{fytas_BC}. In the first-order transition regime, we identified key characteristics indicative of a first-order transition by examining the probability density function of the crystal-field energy and computing the surface tension of the transition using a method proposed by Lee and Kosterlitz~\cite{lee90,lee91}. Our data indicate a linear increase in surface tension as the temperature decreases within the first-order transition regime. This finding is consistent with earlier analysis of the square-lattice model~\cite{kim17}. Subsequently, we confirmed that the triangular Blume-Capel model falls within the universality class of the two-dimensional Ising ferromagnet in the second-order transition regime. This is characterized by the critical exponents $\nu = 1$ and $\gamma/\nu = 7/4$ for the correlation length and magnetic susceptibility, respectively, along with a logarithmic scaling of the specific heat.

Closing, we propose an open research challenge: exploring the phase diagram of the triangular Blume-Capel model using the transfer matrix method with large strip widths enhanced by advanced Monte Carlo techniques and dedicated simulation platforms. This approach would enable the direct estimation of the phase diagram across multiple directions. It would provide valuable insights into the first-order transition regime, leveraging the exact and deterministic nature of the method. Additionally, it would facilitate the identification of the tricritical point and allow for a thorough examination of the second-order regime using phenomenological finite-size scaling based on the correlation length. Such an endeavor would not only complement our current findings but also build upon the successful work of Jung and Kim, who applied this method to the square-lattice Blume-Capel model~\cite{kim17}.

\begin{acknowledgments}
N.G.F. would like to thank Walter Selke, who sadly passed away in December 2023, for his ongoing collaboration on the Blume-Capel model. We are grateful to Nigel Wilding for some very useful comments and suggestions at the final stage of this work. Part of the numerical calculations reported in this paper were performed at the High-Performance Computing cluster CERES of the University of Essex. The work of A.V. and N.G.F. was supported by the  Engineering and Physical Sciences Research Council (grant EP/X026116/1 is acknowledged). D.H.K. acknowledges the support from GIST Research Institute (GRI) grant funded by the GIST and appreciates APCTP and KIAS for their hospitality during the completion of this work.
\end{acknowledgments}

\bibliography{biblio}

\begin{thebibliography}{85}
\bibitem [{com()}]{comment2}%
  \BibitemOpen
  \href@noop {} {}\bibinfo {note} {{Other values of $\omega$ have been reported
  for certain quantities in the two-dimensional Ising model, most notably
  $\omega = 4/3$ and $\omega = 2$. What is more, and to the best of our
  knowledge, for the case of the Blume-Capel model there exists no direct
  numerical or analytical computation of the corrections-to-scaling exponent.
  Fortunately, in our case the observed corrections are rather weak, so that
  numerically we hardly detect any difference between these
  choices.}}\BibitemShut {Stop}%
\end{thebibliography}

\end{document}